\documentstyle[11pt,mrs2001,epsfig]{article}
\begin{document}
\title{STATISTICAL PROPERTIES OF LYMAN BREAK GALAXIES AT z$\sim$4 }

\author{M. OUCHI$^1$, K. SHIMASAKU$^{1,2}$, S. OKAMURA$^{1,2}$, 
	M. DOI$^{1,3}$, H. FURUSAWA$^1$, \\
	M. HAMABE$^4$, M. KIMURA$^5$, Y. KOMIYAMA$^6$, M. MIYAZAKI$^1$,
	S. MIYAZAKI$^7$, \\
	F. NAKATA$^1$, M. SEKIGUCHI$^5$,
	M. YAGI$^7$, and N. YASUDA$^7$}
%SUPRIME-CAM INSTRUMENT TEAM }
\small
\affil{$^1$Department of Astronomy, School of Science, University of Tokyo, Tokyo 113-0033, Japan}
\affil{$^2$Reserch center for the Early Universe, University of Tokyo, Tokyo 113-0033, Japan}  
\affil{$^3$Institute of Astronomy, School of Science, 
        University of Tokyo, Tokyo 181-0015, Japan}
\affil{$^4$Department of Mathematical and Physical Sciences,
        Japan Women's University, Tokyo, Japan}
\affil{$^5$Institute for Cosmic Ray Research, 
        University of Tokyo, Kashiwa, Chiba 277-8582}
\affil{$^6$Subaru Telescope, National Astronomical Observatory, 
        650 N.A'ohoku Place, Hilo, USA}
\affil{$^7$National Astronomical Observatory, 
        Mitaka, Tokyo 181-8588, Japan}
\normalsize

\begin{abstract}
We study the luminosity function and the correlation function of
about 1200 z$\sim$4 Lyman break galaxies (LBGs) with $i'<26$
that are photometrically selected from deep $BRi'$ imaging data
of a 618 arcmin$^2$ area in the Subaru/XMM-Newton Deep Field
taken with Subaru Prime Focus Camera. The contamination and completeness
of our LBG sample are evaluated, on the basis of the 
Hubble Deep Field-North (HDF-N) objects, to be 17\% and 45\%, respectively.
We derive the UV (rest 1700\AA) luminosity functions (LFs) and find
a large population of UV-luminous galaxies at $z\sim4$.
The LFs of the red and blue subsamples imply 
that the bright LBGs are redder in the UV continuum
than the average color of the LBGs.
Then we calculate
the correlation function over $\theta = 2''-1000''$ and
find that it is fitted fairly well by a power law,
$\omega (\theta)=A_\omega \theta^{-0.8}$, 
with $A_\omega = 0.71 \pm 0.26$. We estimate the
correlation length $r_0$ (in comoving units) of
the two-point spatial correlation function 
$\xi(r) = (r/r_0)^{-1.8}$ to be
$r_0=2.7^{+0.5}_{-0.6}$ $h^{-1}$Mpc 
%in a $\Lambda$-dominated universe
($\Omega_m=0.3$ and $\Omega_\Lambda=0.7$).
The correlation function shows an excess of $\omega (\theta)$
on small scales ($\theta \la 5''$), departing from the power-law
fit at $> 3 \sigma$ significance level. Interpreting this as 
being due to galaxy mergers, we evaluate the fraction of
galaxies undergoing mergers to be $3.0\pm0.9\%$, 
which is significantly smaller than those of
galaxies at intermediate redshifts.
\end{abstract}

\section{Introduction}
 In the past several years, deep field surveys have made 
remarkable breakthroughs
in studies on high redshift galaxies. The most successful survey would be 
the Hubble Deep Fields which are deep enough to study
high-{\it z} galaxies up to $z\simeq6$. Many efforts on spectroscopic 
and photometric follow-up observations reveal hundreds of high-{\it z} 
galaxies and their properties (e.g.\cite{fernandez-soto1999}). 
However the surveyed areas of 
HDFs are as small as a few arcmin$^2$ each, so they may be affected
by field-to-field variations.
 Steidel and his collaborators have pioneered in 
statistical studies of high-{\it z} galaxies based on 
ground-based wide survey data \cite{steidel1996}.
They isolate high-{\it z} galaxies
in a two-color plane using the UV continuum features,
and galaxies selected in this way
are called Lyman break galaxies (LBGs).
They studied the number density \cite{steidel1999},
the spatial distribution \cite{giavalisco2001},
nebular emission lines \cite{pettini2001}, and
the stellar population \cite{papovich2001}.
Most of their studies are based on $z\sim3$ LBG samples
which were selected using $U_nGR$ colors.
 In this contribution, we extend the study to
$z\sim4$ LBGs based on our $BRi'$ data taken with a newly installed
Subaru Prime Focus Camera (Suprime-Cam:\cite{miyazaki1998}), 
which is a unique wide-field imager mounted on the 8m Subaru Telescope.
%The data are deep and wide enough for the robust 
%statistical studies of z=4 LBGs.
Throughout this contribution, magnitudes are in the AB system,
and all calculations assume a $\Lambda$-dominated 
spatially flat cosmology, 
$(\Omega_m,\Omega_\Lambda)=(0.3, 0.7)$.

\section{Observations and Data Reduction}
Deep and wide-field $B$-,$V$-,$R$-, and $i'$-band
imaging data of a central $30' \times 24'$ area
in the Subaru/XMM-Newton Deep Survey Field 
($2^h 18^m 00^s$,$-5^\circ 12 ' 00''$[J2000])
were taken with Suprime-Cam
during the commissioning observing runs
on 2000 November 24-27. The present work is
based on the $B$,$R$, and $i'$ data. The individual CCD data
were reduced and combined using IRAF and the mosaic-CCD data
reduction software developed by us \cite{yagi1998}. The final 
images cover a contiguous 618 arcmin$^2$ area with a point-spread function
FWHM of $0''.98$. The net exposure times of the final images
are 177, 58, and 45 minutes for $B$, $R$, and $i'$, respectively.
The limiting magnitudes are $B=27.6$, $R=26.5$, and $i'=26.2$
for a $3 \sigma$ detection in a $2''$ diameter aperture. 
The $i'$-band frame is chosen to detect objects,
and we limit the object catalog to $i'\la26$, in
order to provide a reasonable level of photometric 
completeness.

\section{Selection of z$\sim$4 Lyman Break Galaxies}
\label{sec:lbgselection}
\begin{figure}
\plotone{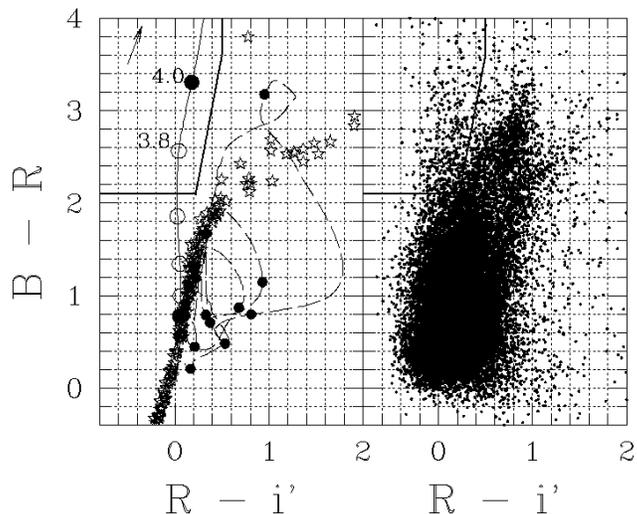}{8.25cm}
%\plotone{images/BRi_diagram_disp_marseille2.eps}{7.3cm}
\caption{{\it Left panel}: $B-R$ vs. $R-i'$ color diagram displaying
the colors of model galaxies and stars. 
Typical spectra of elliptical, Sbc, Scd and irregular galaxies
\cite{coleman1980} are redshifted from $z=0$ to $z=3$,
which are shown by four dashed lines.
Each line has filled circles at $z$=0,1,2 and 3.
We use a typical spectrum of galaxies 
at $z\simeq 3$ to compute the $z\sim4$ galaxy track 
which is shown by the solid line \cite{ouchi2001a}.
Star marks are 175 Galactic stars given by \cite{gunn1983}.
The arrow expresses the direction and magnitude of dust
extinction on the $B-R$ vs. $R-i'$ plane for $E(B-V)=0.15$, 
when Calzetti's extinction law \cite{calzetti1997} is applied.
The box surrounding the upper left region is the selection
criteria of our $z\sim 4$ galaxies (equation \ref{eq:lbgselection}).
No obvious contaminant is found within the box. 
{\it Right panel}: $B-R$ vs. $R-i'$ color diagram displaying
the colors of 5 $\sigma$ detected objects in our data.
The box for our selection criteria is also shown. 
}
\label{fig:BRi_diagram_disp}
\end{figure}
 Our catalog contains 42,557 objects with $i'\leq 26.0$
in total. On the basis of expectations from GISSEL96
\cite{bruzual1993} population synthesis models,
%\footnote{
% The model parameters are chosen to match the
%observed colors of $z=3$ galaxies: the Salpeter initial 
%mass function, $Z_{metal}=0.2Z_{\sun}$, and an age of
%10 Myr for an instantanious burst (\cite{sawicki1998}).
%}
%with dust attenuation of $E(B-V)=0-0.5$, assuming 
%Calzetti's extinction law (\cite{calzetti1997}),
we define the photometric selection criteria for
galaxies at $z\sim4$ (\cite{ouchi2001a} for model parameters) as
\begin{equation}
B-R > 2.1, 
R-i' < 0.5, 
B-R > 5.4(R-i')+0.9 \\
\label{eq:lbgselection}
\end{equation}
We estimate the redshift distribution, $N(z)$, of 
galaxies satisfying equation (\ref{eq:lbgselection})
from the HDF-N objects 
for which magnitudes,
colors, and photometric redshifts are given 
in \cite{fernandez-soto1999}. We find that 
the criteria select $z=3.8\pm 0.5$ galaxies.
%
%We estimate from the models that the expected redshift
%distribution of galaxies satisfying 
%equation\ref{eq:lbgselection} is 
%$z=3.8\pm 0.5$(\cite{ouchi2001b}).
%
 There are a total of 1192 objects that meet the criteria.
Figure \ref{fig:BRi_diagram_disp} shows the $B-R$ vs. $R-i'$
color diagram for model galaxies and
Gunn \& Stryker's stars \cite{gunn1983} ({\it left panel}), and
for $5 \sigma$ detected objects in our data ({\it right panel}).
The left panel demonstrates
that $z\sim4$ galaxies are well isolated from interlopers, i.e., 
low-{\it z} galaxies and Galactic stars.
 We have estimated the contamination and completeness of 
our LBG sample by Monte Carlo simulations, generating artificial
objects which mimic the HDF-N galaxies 
%for which magnitudes,
%colors, and photometric redshifts are given in \cite{fernandez-soto1999}
and distributing them on our original images. 
The contamination is defined, for the detected simulated objects,
as the ratio of low-redshift ($z<3.3$) objects meeting equation
(\ref{eq:lbgselection}) to all the objects satisfying equation 
(\ref{eq:lbgselection}). The completeness is defined as
the ratio of $z\simeq4$ simulated objects passing our detection
threshold $and$ satisfying equation (\ref{eq:lbgselection}) to
all (detected $+$ undetected) $z\simeq4$ simulated objects. We
find from the simulations that the completeness and the contamination
are 45\% and 17\%, respectively.
%
%The contamination,
%$f(cont)$, and the completeness, $f(comp)$, are defined
%for the simulated objects as 
%
%\begin{eqnarray}
%f(cont)=\frac{N[detected low-redshift(z<3.3) objects meeting eq.(\ref{eq:lbgselection})]}{N[detected objects meeting eq.(\ref{eq:lbgselection})]}\\
%f(comp)=\frac{N[detected z\sim 4 objects meeting eq.(\ref{eq:lbgselection})]}{N[detected and undetected z\sim 4 objects]} \ \ \ \ \ \ \ \ \ \ \ \ \ \ \ \ \ \ \ \ \ \ \
%\end{eqnarray}
%
% From the simulations, we find $f(cont)=17\%$ and 
%$f(comp)=45\%$, respectively. 
%
%Note that the completeness 
%defined here is a combination of the detection
%completeness and the completeness of color selection.
%
%Our clor criteria effectively sample HDF-N galaxies at 
%$3.3 < z < 4.3$ with a selection completeness of 89\%.
%Namely, 89\% of $z\simeq$4 galaxies pass equation (\ref{eq:lbgselection}) once
%they are detected. We adopt for the redshift distribution
%$N(z)$ of LBGs in our sample the one derived from the simulations
%takign account of the contamination and completeness.

\section{Luminosity Function}
\label{sec:lumifun}

\begin{figure}
\plottwo{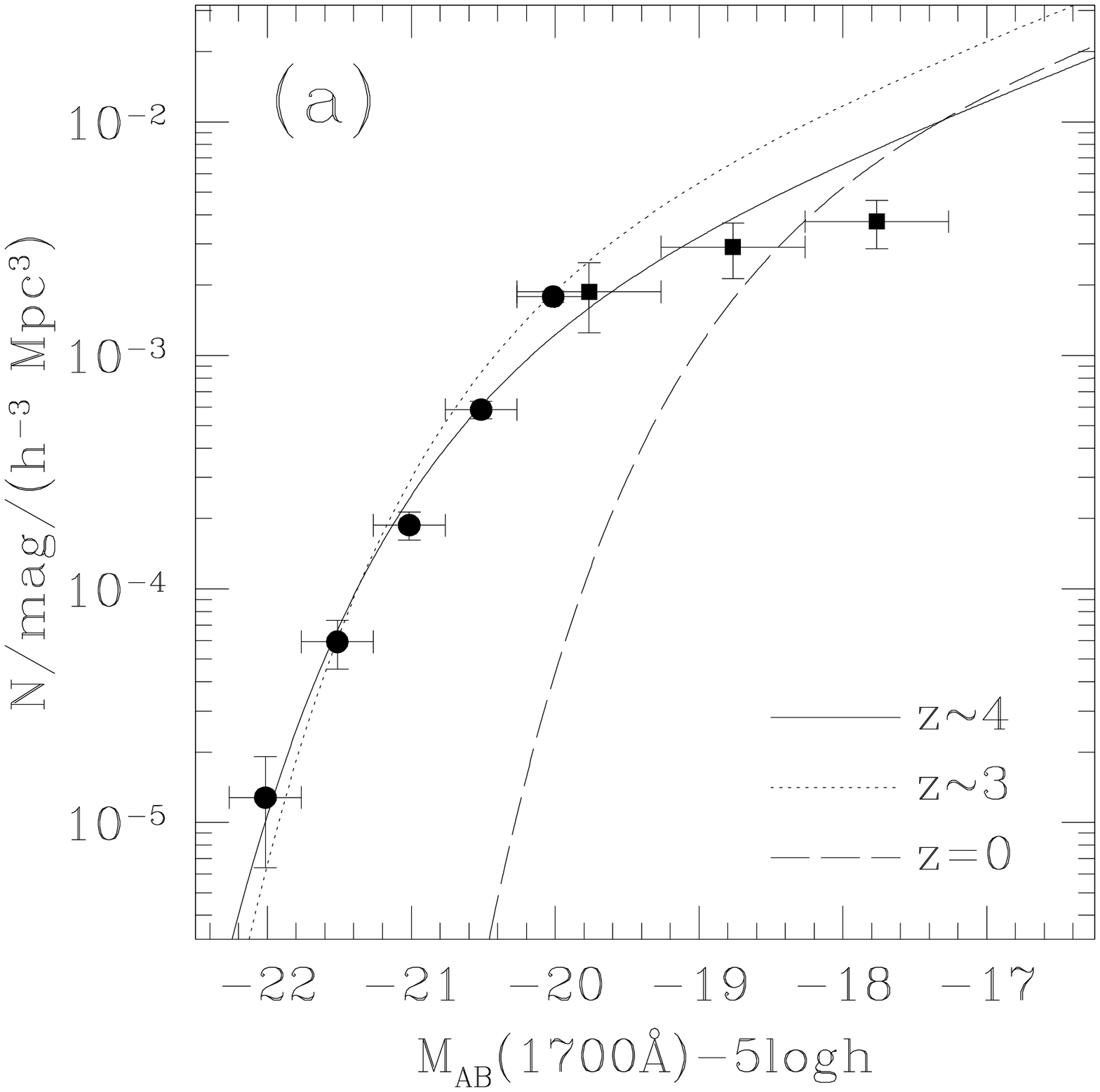}{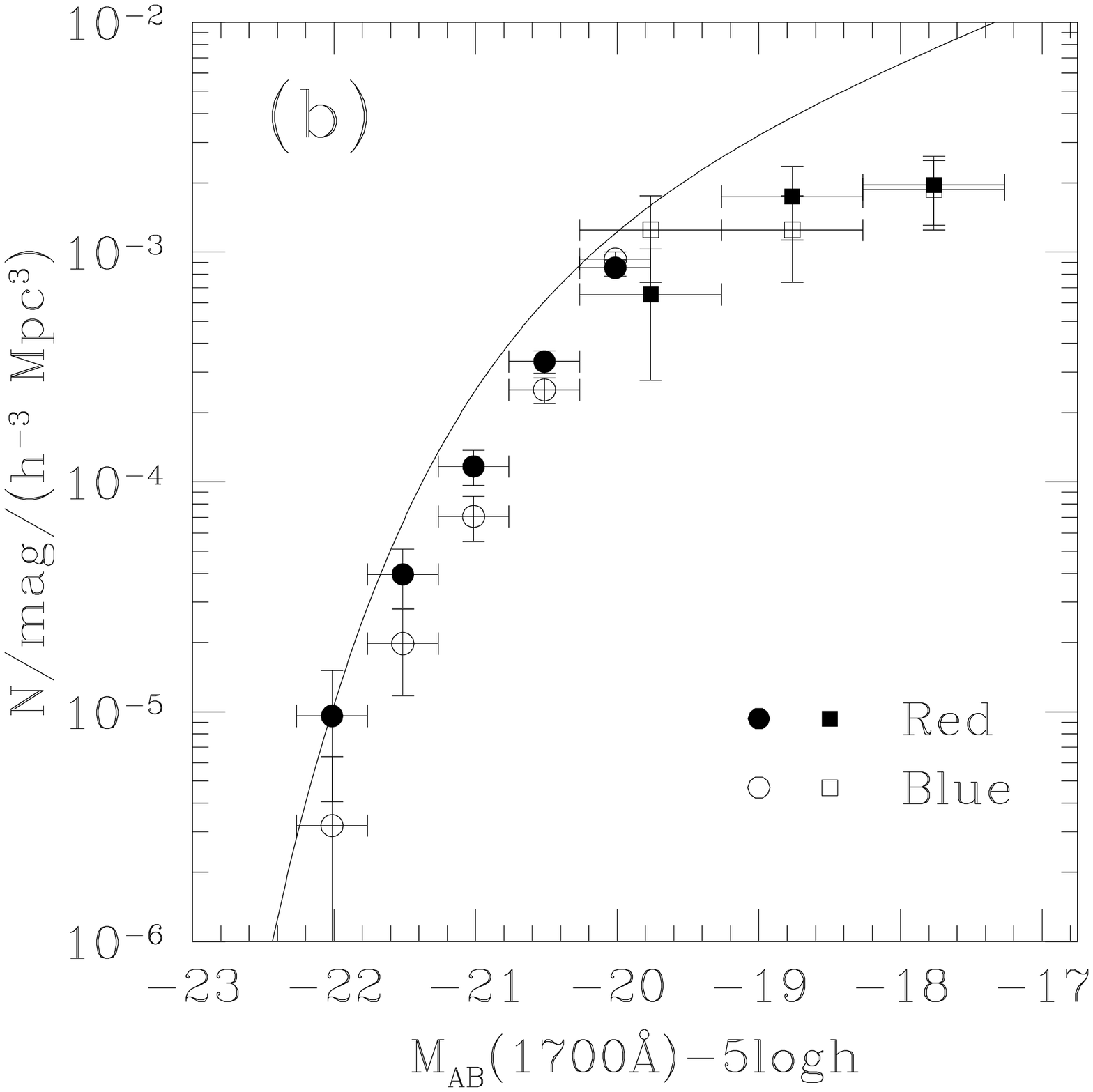}
%\plottwo{images/lumifun_marseille2001_2.eps}{images/lumifun_marseille2001_color_2.eps}
\caption{(a) Luminosity function (LF) of $z\sim4$ LBGs, together with
those for $z=0$ and $3$ galaxies. 
The filled circles indicate $z\sim4$ 
LBGs derived from our data, and the squares are for the $z\sim4$ galaxies
from the 
HDF photo-{\it z} catalog \cite{fernandez-soto1999}. The solid line shows 
the best fitted Schechter function to the $z\sim4$ data.
The dashed line is the LF of UV selected galaxies at $z=0$ (whose
magnitudes are in $M_{AB}(2000$\AA$)$ \cite{sullivan2000}.) 
The dotted line denotes the LF of $z\sim3$ LBGs \cite{steidel1999}.
A large population of UV-luminous ($M_{AB}\la -20$) galaxies is seen 
at $z\sim3$ and 4. 
(b) LFs of the red and blue subsamples. The filled circles and
filled squares show the red sample, while the open circles
and open squares are for the blue sample. The solid line is 
the best fitted Schechter function obtained from the whole 
sample, which is shown as the solid line in (a). See text for details.
}
\label{fig:lumifun}
\end{figure}

 The UV luminosity function (LF) of $z\sim4$ LBGs is derived from
our 1192 objects with $i'<26$.
We calculate the surveyed effective 
volume which is the integrated volume 
from $z=3.3$ to $z=4.3$ 
by taking into account the
completeness and contamination of the sample
selection, and N(z) (\S \ref{sec:lbgselection}).
%We calculate the surveyed effective 
%volume by integrating $dV_{eff}/dz$ from $z=3.3$ to $z=4.3$.
The LF is shown in figure \ref{fig:lumifun}(a), 
together with those at $z=0$
\cite{sullivan2000} and $z\sim3$ \cite{steidel1999}.
The LF of our $z\sim4$ LBG sample is 
consistent with the one derived by \cite{steidel1999}.
We fit the Schechter function to the LF, and find
$M^*_{AB}(1700$\AA$)=-20.41$, $\phi^*=1.9\times10^{-3}$ 
with a fixed slope $\alpha=-1.6$. 
It is found from figure \ref{fig:lumifun}(a), that 
all the LFs from z=0 to 4 seem to prefer a steep slope, 
$\alpha\simeq -1.6$.
The $M^*$s at $z\sim$3 and 4 are about 2 magnitude 
brighter than that at z=0. The number of
bright ($M_{AB}(1700\AA)\la -20$) galaxies are
much larger at high redshifts than in the local
universe, while that of faint galaxies are 
comparable. This implies that
a large population of UV-luminous 
galaxies exists only at high redshifts.
%
%\footnote{We calculate the LFs
%assuming the other cosmologies,
%$(\Omega_m,\Omega_\Lambda)=(1.0, 0.0)$
%$(0.3, 0.0)$, this UV-luminous population
%is also found at a high significance level.}
%
This population reflects the 
high cosmic starformation rates 
at high redshifts, which were derived by 
\cite{madau1996} and \cite{steidel1999}. 
There are two extreme explanations for this population.
One is 
that the star-formation rate of $z\sim3-4$ galaxies
is intrinsically higher than that for $z=0$ galaxies,
and the other is that the $z\sim3-4$ galaxies are very 
dust-poor (or partially dust free)
and thus the $M^*$ values are apparently brighter
than that for $z=0$ galaxies which suffer from non-neglibible
dust extinction.
%
%the existance of very active star-forming
%galaxies, another is the existance of very dust 
%poor ( or partially dust free ) galaxies. 
A detailed discussion will be given in \cite{ouchi2001b}.
%
% phi_star = 0.00193244
% m_star = -20.31221-0.1
% alpha = -1.6
%
% phi_star_sullivan = 0.00912549
% m_star_sullivan = -18.3498

 We divide our whole LBG sample into two subsamples,
{\it blue} and {\it red}, by the $R-i'$ color. 
The blue (red) sample is composed of objects
bluer (redder) than the median color of the 
whole sample, $\left< R-i'\right>=0.12$.
The LFs of red and blue samples are calculated 
in the same manner
as for the whole sample. Figure \ref{fig:lumifun}(b)
shows the LFs of red and blue samples. These LFs are not
significantly different at the faint end ($M_{AB}>-20.3$), however,
the LF of red sample seems to be higher than that of the blue 
sample at the bright end ($M_{AB}<-20.3$). This may imply that
the bright LBGs are dominated by galaxies red 
in the $R-i'$ color.  The $R-i$ color
reflects the shape of UV continuum between
$1380-1620$\AA \ for $z=3.8$ LBGs 
\footnote{ $1530$\AA$-1810$\AA \ for $z=3.3$ LBGs, 
$1250$\AA$-1470$\AA \ for $z=4.3$ LBGs}, 
and the difference in the 
color is thought to be caused by various reasons,
for example, differences in dust extinction, 
stellar population, and redshift. 
If the difference in the color
is mainly originated from the dust extinction, the 
boundary of the color, $R-i=0.12$, is estimated to
be $E(B-V)\simeq 0.18$, following \cite{meurer1999}.
If this is the case, the brighter LBGs may be dominated 
by dusty LBGs with $E(B-V)> 0.18$. Note that
in deriving the red and blue LFs, we assume that the
completeness and the contamination of these subsamples
are the same as for the whole sample.
We should point out the possibility that 
the difference in red and blue LFs comes from
a systematic difference in the completeness 
and contamination factors of the blue and red samples.

\section{Angular Correlation Function}

\begin{figure}
\plottwo{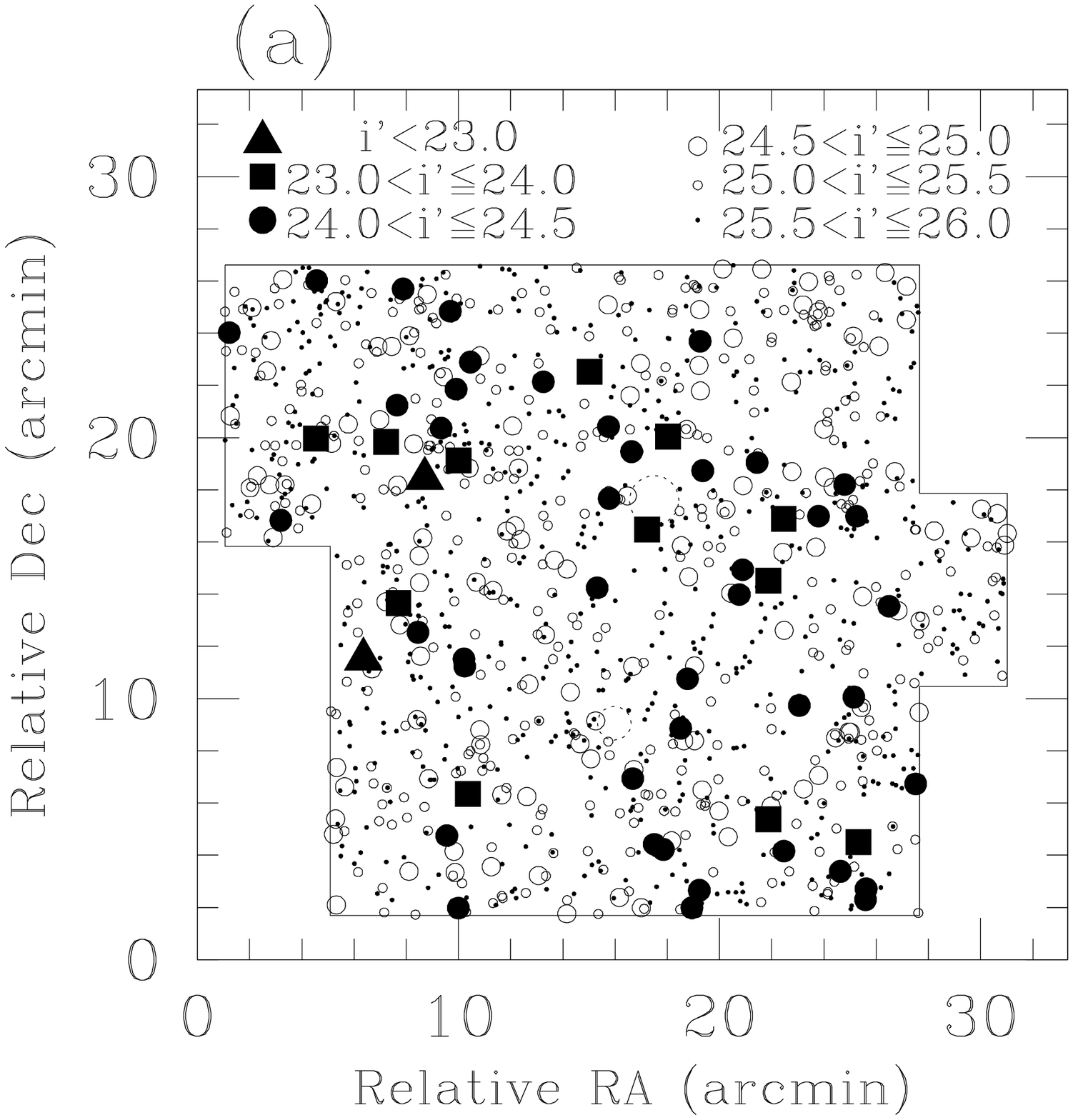}{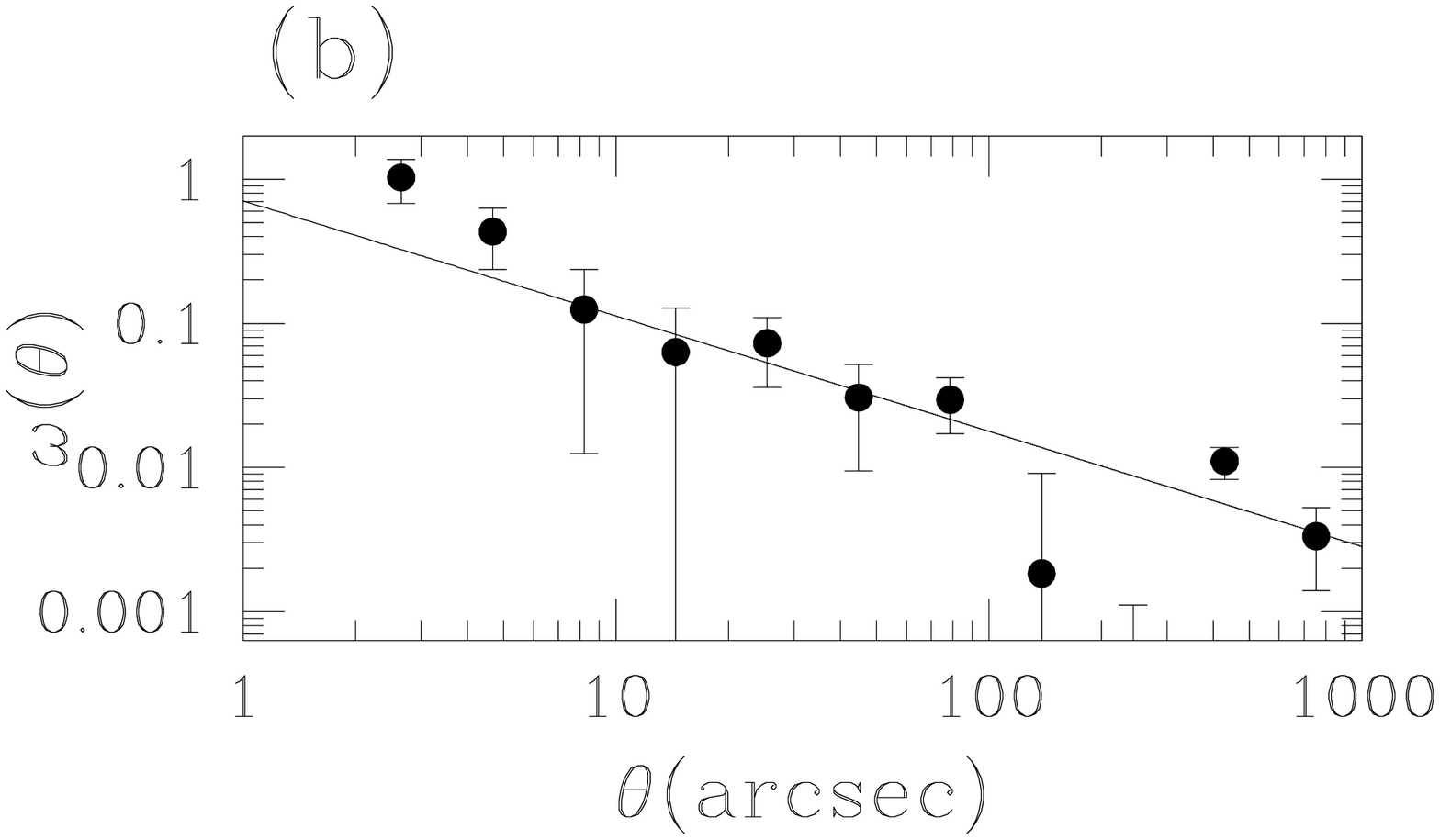}
%\plottwo{images/lbgdistribution_marseille2001.eps}{images/resultacorr_all_marseille2001.eps}
\caption{(a) Sky distribution of the $z\sim4$ LBGs.
Different symbols correspond to different magnitude bins defined 
in the panel. Masked regions, to avoid the effects of bright stars, are
shown as dotted circles.
(b) Angular correlation function for 
the $z\sim4$ LBGs. The solid line shows the best-fit power law with 
$\omega(\theta)=A_\omega \theta^{-0.8}$. }
\label{fig:corr}
\end{figure}

 Figure \ref{fig:corr}(a) shows the 
sky distribtuion of $z\sim4$ LBGs in our sample.
In this figure, we find a somewhat inhomogeneous
distribtution of LBGs, especially for bright
($i'<24.5$) ones.
 We derive the angular two-point correlation function
$\omega(\theta)$ using the estimator defined by 
\cite{landy1993} (see \cite{ouchi2001a}).
%$
%\omega_{obs}(\theta)
%  = [DD(\theta)-2DR(\theta)+RR(\theta)]/RR(\theta),
%$
%where $DD(\theta)$, $DR(\theta)$, and $RR(\theta)$ are numbers of
%galaxy-galaxy, galaxy-random, and random-random pairs normalized by
%the total number of pairs in the data samples.
%The random sample is composed of 200,000 sources with the same
%geometrical constraints of the data sample. 
%The formal error \footnote{The formal error does not 
%include sample variance, which is
%caused from field-to-field variations.}
% in $\omega$($\theta$) is assigned by
%$\sigma_\omega = \sqrt{[1+\omega_{obs}(\theta)]/RR}$.
%
%The real correlation function $\omega(\theta)$ is
%offset by the integral constant ($IC$: \cite{groth1997});
%$\omega(\theta)= \omega_{obs}(\theta)+IC$
%,where $IC$ of the sample is calculated to be 0.0045.
%
The resulting angular correlation function
for the sample is shown in 
figure \ref{fig:corr}(b)
\footnote{Error bars do not 
include sample variance caused from 
field-to-field variations. 
The effect of the variations in our sample is 
propably modest since our sample probes a large comoving volume,
$36\times36\times520 = 6.7\times10^5 $h$^{-1}$ Mpc$^3$.}.

%after the 
%application of $IC$. 
We fit a power law, 
$\omega(\theta)=A_\omega \theta^{-\beta}$, 
to the data points, and find the slope 
$\beta = 0.6^{+0.6}_{-0.4}$, which is 
consistent with those for nearby galaxies, $\beta = 0.7-0.8$, 
though the errors in our estimate are large.
The best fit value for $\beta \equiv 0.8$
is $A_\omega=0.71\pm0.26$ in units of arcsec$^{\beta}$.

The two-point angular correlation function is related
to the spatial correlation function 
$\xi(r) = (r/r_0)^{-\gamma}$ by an integral equation,
the Limber transformation \cite{peebles1980}.
We apply the Limber transformation to the best fit value
of $A_\omega$ with $\beta \equiv 0.8$, 
using the redshift distribution $N(z)$ of the LBGs
(\S \ref{sec:lbgselection}). 
The obtained correlation length is $r_0=2.7^{+0.5}_{-0.6} h^{-1}$ Mpc.
%for a $\Lambda$-dominated spatially flat cosmology, 
%$(\Omega_m,\Omega_\Lambda)=(0.3, 0.7)$ 
%( \cite{ouchi2001a}).
%
This is twice as large as the correlation length of the dark matter
at $z\simeq4$ predicted from an analytic model by \cite{peacock1996}.
The difference is thought to be due to the biasing of galaxy
distribution, which is characterized by the biasing factor, $b$,
as $\xi_{gal}(r)=b^2 \xi_{matter}(r)$.
We obtain a linear bias $b \simeq 2.6$, which is
similar to that found by \cite{giavalisco2001}
for bright $z\simeq3$ LBGs. On the other hand, the 
semianalytic models predict much stronger clustering
for galaxies ( eg. \cite{baugh1999}, 
\cite{kauffmann1999}). A further discussion
requires a detailed comparison of galaxy properties
and selection criteria, between
galaxies in the models and LBGs in our sample. 

 In figure \ref{fig:corr}(b), we find an excess
\footnote{With a visual inspection, we find that all the
galaxies contributing to the excess are real; no artificial objects,
such as halos of bright stars, are found to be included in them.}
of $\omega(\theta)$ at small scales ($\theta \la 5''$) at 
$> 3 \sigma$ significance levels, relative to the best-fit power law. 
This can be caused by various effects; for example, 
galaxy-galaxy mergers, two galaxies within a common 
dark matter halo, and/or field variance.
If we interpret this as galaxy-galaxy mergers of those 
$i'<26$ LBGs, which are classified as bright LBGs
with $M(1700\AA) \la M^*(1700\AA)$ (\S \ref{sec:lumifun}),
we can calculate, following \cite{roche1999}, 
the fraction of galaxies undergoing mergers as
from the observed and expected numbers of galaxy pairs 
with a separation of $1\farcs5 < \theta < 4''$ 
(the projected physical separation corresponds to
7.5 to 20 $h^{-1}$kpc at $z=3.8$).
The resulting value is $f_{pair} = 3.0 \pm 0.9\%$.
This value is significantly smaller than 
those for intermediate-redshift ($z \sim 0.3$) galaxies whose 
$f_{pair}$ at $\la $ 20 $h^{-1}$kpc is $5-15\%$  
(e.g. \cite{roche1999}).

%\section{Summary}

%\begin{table}
%\begin{center}
%\caption{my table}
%\begin{tabular}{lll}
%\hline
%X               & Y       & Z         \\ 
%\hline
%$2.3 \times 10^4$ & 2.0 cm  & 2.3 cm    \\
%$9.8 \times 10^5$ & 2.0 cm  & 2.3 cm    \\
%$5.4 \times 10^4$ & 2.4 cm  & 1.5 cm    \\
%$6.3 \times 10^4$ & 2.4 cm  & 1.5 cm    \\
%\hline \hline
%\end{tabular}
%\end{center}
%\end{table}

\acknowledgements{We would like to thank the Subaru
Telescope staff for their invaluable help in commissioning the
Suprime-Cam that makes these difficult observations possible.
M. Ouchi, H. Furusawa, and F. Nakata acknowledges support 
from the Japan Society for the 
Promotion of Science (JSPS) through JSPS Research Fellowships 
for Young Scientists.}

\vfill
\end{document}